\documentstyle[aps,preprint,12pt]{revtex} 
\begin{document} \draft 
\title{Moving a perturbation across quantum dots:  tuning of Fano resonances}

\author{M.  Mendoza and P.  A.  Schulz }

\address{Instituto de F\'{\i}sica Gleb Wataghin, UNICAMP, Cx.P.  6165, 13083-970, Campinas, SP, Brazil}

\maketitle

\date{today}

\begin{abstract} We propose a controllable way of tuning Fano resonances in open quantum dots in the
absence of magnetic field.  A quantum dot can be modified by changing the gate voltages that define the
dot itself.  An extra degree of freedom can be introduced by means of a controllable repulsive
perturbation, such as the one induced by scanning a Atomic Force Microscope tip on a real sample.  We
numerically investigate the coupling between localized states in the dot and the continuum in the
leads.  The advantage of such position controllable perturbation is the selective manipulation of the
quantum dot states.  We show that this could be a feasible alternative to quantum dots in Aharonov-Bohm
interferometers as a Fano resonance tuning device.

\noindent PACS number(s) 73.40.Gk,73.21.Fg,73.21.La \end{abstract}

\newpage


In the past few years, mesoscopic systems became appropriate tools for investigating wave function
interference effects.  The nanometer-scale technology turned into reality structures with dimensions 
that are smaller than the phase coherence length.  The prototype
are systems created in a two-dimensional high-mobility electron gas (2DEG) at the interface of a
GaAs/AlGaAs heterostructure structured by means of the split gates technique \cite{Kouwenhoven}.
 Two important examples should be mentioned here:
   the Quantum Point Contact (QPC), where conductance quantization
has been seen \cite{van Wess}; and the Quantum Dot (QD), with a wide range of
interesting transport properties  
\cite{Lundberg}.  
Starting from these two devices, systems of greater complexity
have been conceived and fabricated for controling  electronic wave function interference effects. 
 A recent example of such
effort is the imaging of the coherent electron flow through QPC, where the interference effects can be
controlled by scanning an Atomic Force Microscope (AFM) tip over the sample 
\cite{Topinka}. 
Another example, related to the present work is the tunable Fano resonance 
 in an Aharonov-Bohm interferometer with an embebed QD 
\cite{Kobayashi}.
We propose that Fano resonances could be completely tunable in open quantum dot (OQD) structures  by moving 
a local perturbation across the system, like a tip of an AFM. 
Indeed such a procedure enables a selective manipulation of the QD states. 
Up to now, in absence of magnetic fields, Fano 
resonances have been only partially tuned by changing gate voltages in single-electron transistors. 
\cite {Gores}.

 In the present approach, a continuous system is discretized into a
 lattice, considering a single $s$-like orbital per site and only nearest-neighbour hopping elements.
 The OQD structure, emulated by a tight-binding lattice model is depicted in Fig.1(a).  The black
 circle represent a potential columm simulating the perturbation induced, for instance, by an AFM tip
 located on the sample at that position.  In what follows we consider perturbations of a single host
 lattice site ($S_p$), which corresponds to a extension relative to the quantum dot of
 $S_p$/$S_{QD}$=1/49 for OQD with dimensions $Lx=Ly=7a$, where $a$ is the host lattice parameter
 (Fig.1(a)). The repulsive perturbation will be characterized by a height $H$ in the 
 range between $H=50$ meV and $H\approx1$ eV.
  The tight-binding hopping parameters are chosen in order to emulate the electronic
 effective mass for the GaAs bottom of the conduction band, $m^*=0.067m_0$.  Hence,
 $V_{x,y}=-\hbar^2/(2m^*a^2)$, $V_{x,y}=-0.142$ eV for a lattice parameter of $a=20$ \AA. 
 Such parametrization represents quantum dots with lateral sizes up to 
$L_x = L_y =140 \AA$, at least
one order of 
magnitude lower than the typical dimensions of actual quantum dots constructed 
by litographic methods. 
However, the present work has the proposal of illustrating a selective manipulation of 
confined states. Here the relevant energy scale is $\Delta E \approx 0.1$ eV. For systems ten times 
larger, the new energy scale would be of the order of $1$ meV, still accessible for available temperatures. 
Another relevant scale is the ratio between the extension of the perturbative 
spike and the dot dimension, $S_p/S_{QD}$. A previous work 
\cite{Mendoza} 
suggests that the energy shifts are qualitatively insensitive to this ratio if the perturbation height, $H$, 
is less than $\Delta E $.
  
The AFM tip can also be seen as a controllable impurity in a quantum dot and therefore a simple tunable
experimental realization of a multiply connected nanostructure 
\cite {Joe}.
 
We will interested in the transmission probabilities and local density of states (LDOS), which are obtained 
by means of Green's functions approaches. For the LDOS we follow a self-energy method 
\cite {Datta}, 
the total Hamiltonian, $H_T$, is a sum of four terms:  the dot connected to the two point contacts 
regions, described by the $H_D$, the left and rigth contact regions, $H_L$ and $H_R$, respectively; as well 
as the coupling term between the contacts and the dot structure, $V$:

\begin{equation} H_T=H_D+H_L+H_R+V= \left(\begin{array}{ccc} H_L&\tau_L&0\\ \tau_L^+&H_D&\tau_R^+\\
0&\tau_R&H_R \end{array} \right) \end{equation}

where $\tau_{L(R)}$ describes the interaction between device and contacts.  The corresponding Green's
function is given by

\begin{equation} (E-H_T)G_T(E)=I \end{equation}

We define the Green's functions for the isolated contacts:

\begin{eqnarray} (E-H_L)g_L(E)=I&\mbox{ and }&(E-H_R)g_R(E)=I \end{eqnarray}

Using equation (1) in (2) we obtain:

\begin{equation} \left(\begin{array}{ccc} E-H_L&-\tau_L&0\\ -\tau_L^+&E-H_D&-\tau_R^+\\ 0&-\tau_R&E-H_R
\end{array} \right) \left(\begin{array}{ccc} G_L&G_{LD}&G_{LR}\\ G_{DL}&G_D&G_{DR}\\ G_{RL}&G_{RD}&G_R
\end{array} \right) = \left(\begin{array}{ccc} I&0&0\\ 0&I&0\\ 0&0&I \end{array} \right) \end{equation}

The Green`s function of the device, $G_D$, result in

\begin{equation} G_D=(E-H_D-\Sigma^L-\Sigma^R)^{-1} \end{equation}

where $\Sigma^L=\tau_L^+g_L\tau_L$ and $\Sigma^R=\tau_R^+g_R\tau_R$ are the so called self-energies
which describe the effects of the leads on the device.  The propagators in the transverse mode
representation for the isolated contacts, $\tilde g_{L(R)}$ are:

\begin{equation} <\nu^{'}|\tilde g_{L(R)}|\nu>=\frac{e^{i\theta_{\nu}}}{|V_x|}\delta_{\nu^{'},\nu}
\end{equation}

with

\begin{equation} \theta_{\nu}=cos^{-1}[\frac{(E-\epsilon _{\nu})}{2V_x}+1] \end{equation}

The matrix elements of the unitary transformation matrix, $U$, which transforms the Green`s function
from site to transverse mode representation ($\tilde g_{L(R)}=U^+g_{L(R)}U$) are:

\begin{equation} <n|U|\nu>=\chi _{n}^{\nu}=\sqrt{\frac{2}{N+1}}sin(\frac{\pi \nu n}{N+1})
\end{equation}

where n (transverse site) and $\nu$ (transverse mode) are row and columm indexes of the $U$ matrix, 
respectively.    The $g_{L(R)}$, $\tau_{L(R)}$, $\Sigma^{L(R)}$ and $U$ are $N$x$N$
matrices, where $N$
is the number of transverse sites.  Using equation (5) we obtain  the 
LDOS:

\begin{equation} \rho_E(m,n)=\frac{-1}{\pi}Im[G_D(m,n;E)] \end{equation}

where m is the longitudinal site.  This LDOS is proportional to the probability density of
a state of energy E.  Finally, a Green's function is related to the conductance  by the Landauer-B\"uttiker 
formula 
\cite{Landauer}, 
by means of a recursive solution of the Dyson equation as  discussed elsewhere 
\cite{ferry}.


 We can control the coupling of  OQD bound
states to the two dimensional reservoirs by changing the width, $w$, of the left and right QPCs, 
Fig. 1(a)
 Transmission probabilities as a function of the incident electron energy are shown in Fig. 1(b).
   For $w=3a$,
 (continuous line) we clearly see resonances in the transmission probability due to the bound states of
the OQD, as well as the first conductance plateau.  For wider leads, $w=5a$, (dashed line) we see the
resonances due to OQD states shifted to lower energies, since the effective confinement is diminished
with the widening of the contacts.  Moreover the resonance peaks are broadened because the coupling to
the reservoirs is enhanced.  As expected, the conductance plateaus are also shifted to lower energies
and the onset of a second plateau can be seen in the same energy range.  Two aspects related to
widening the leads should be kept in mind; (i) the shift in energy and broadening of the resonances are
not a monotonous function of the energy; and (ii), even the shape of a resonance may the qualitatively
changed, as can be seen for the antiresonance for $w=3a$ changing into a peak for $w=5a$.

The previous paragraph summarizes the effects on the conductance that can be achieved by varying
(symmetrically) the leads, i.e, the voltages of the gate electrodes that actually define the quantum
dot structure.  In what follows we will be interested in effects due to a new degree of freedom in the
manipulation of the dot states, as well as the coupling between them and the continua of states in the
leads.  Such new mechanism is introduced by a local potential perturbation that can be moved across the
sample, like an AFM tip, as suggested above.  The novelty of this local perturbation is that the
resonances may be selectively tuned, exploring the wave function symmetries of the states.

In order to explore the wave function symmetries, one has to look systematically along three symmetry
axis for the present case of a square QD:  along the axis connecting the leads ( x axis ), the axis
perpendicular to the line joining the leads and crossing the center of the dot ( y axis ), and a
diagonal, also crossing the center, as indicated in the sketch shown in Fig.  1(a).  The shift in energy
of a certain resonance peak depends on the position of the local perturbation, since such shift is
proportional to the wave function amplitude at this position \cite{Salis}:

\begin{equation} E_i^{'}=E_i+V|\psi(x,y)|^2 \end{equation}

where $E_i$, is the unperturbed energy of state $i$, $V$ is the effective perturbative potential, and
$|\psi(x,y)|^2$ is the probability density of the unperturbed state.  
Hence, a mapping of the probability density related
to a resonance peak can be obtained by a contour plot of the energy shifts as a function of the
perturbation position 
\cite{Mendoza}.  
Here we are interested not necessarily in wave function mapping,
but in the related selective tuning of the resonances.  Fig.  1(c) shows contour plots of the LDOS at
the energies of the 4 resonances and the anti-resonance shown in Fig.  1(b) for the $w=3a$ wide leads
case (continuous line).  The last pannel is for an energy sligthly above the anti-resonance.  Having
these contour plots in mind it is straightforward to see that the width of the resonance peak is
proportional to the wave function amplitude (LDOS) leaking into the leads.  Of particular interest is the
strong localization of the fifth state, which correspond to the anti-resonance in Fig.  1(b), as well
as the clear 1d channel character at off resonance energies in the conductance plateau, $E_5+\Delta$,
Fig.  1(c).

Within this framework illustrated by Fig. 1 
we are able to understand the effects of the position dependent
perturbation.  In this sense, two important effects are shown in Fig.2.  First, for a perturbative
spike positioned at the center of the quantum dot, Fig.2(a), we see the selective shift of the
resonances in the perturbed system with $H=0.2$ eV(continuous line) 
compared to the unperturbed one (dashed line).  We
clearly identify that important shifts (for the lowest resonance and the anti-resonance) occur only for
those states with significant amplitude at the center of the dot.  Next, for a perturbation $H= 0.5$ eV 
placed at
one of the leads (continuous line), Fig. 2(b), no energy shifts of the resonances 
are observed.  On the other hand, a strong suppression of the conductance plateau
can be observed, an effect which has been used recently for imaging the electron flow through quantum
point contacts 
\cite{Topinka}.

The dependence of the selective resonance shifts on the wave function symmetries for moving the
perturbative spike along the y axis can be analyzed by comparying Fig. 2(a) and Fig. 2(c).  Now, for 
$H= 0.05$ eV (continuous line) and $H=4$eV (dashed line),  the
second resonance remains almost unchanged and the lowest one is shifted, as expected from the LDOS in 
Fig. 1(c).
Having the LDOS at the resonance energies in mind, it is also expected that the third resonance would be
perturbed in this situation and, indeed, this resonance is not only shifted in energy but the line
shape changes into a Fano-like form.  Here the shift in energy affects dramatically the coupling with
the continua in the leads and therefore the line shape change in the transmission probability.  Moving
the perturbation further away from the center in the y direction, Fig.  2(d), this third resonance
evolves towards a symmetric peak again ( continuous line for $H=0.05$eV),
illustrating the tunability of a specific resonance with a controllable perturbation.

The symmetry of the localized states in the QD give rise to two new consequences when the 
perturbation is moved along the diagonal, Fig. 3. 
Here dashed and continuous line are for perturbative spikes at the center and at the diagonal, respectively.
Both situations are for $H=0.5$ eV. First, we observe the inversion of the Fano resonance associated 
to $E_3$ \cite {kim}: from Fig. 2 (c) to Fig. 3, the peak position is reversed. 
 Moreover,
only an off center perturbation along the diagonal breaks the symmetry of the $E_4$ state, 
allowing the coupling with the 1D channel and leading to a Fano resonance. This symmetry is not 
broken with the perturbation positioned along the other symmetry axis (x and y), since the LDOS 
related to $E_4$ is negligible there.

In summary, we propose that scanning an AFM across an open quantum dot could be an efficient tool for 
a selective manipulation of quantum states. The local perturbation induced by the tip is an important 
new degree of freedom in order to explore and modify resonant states. 
As an example, we show that the coupling of a 
specific level with the continuum can be tuned over a wide range in which the line shape of the associated 
resonance in the conductance changes from a symmetric peak into a asymmetric Fano resonance, which can be 
inverted or tuned back into a single peak. This effect suggests an interesting alternative to quantum 
dots in Aharonov-Bohm interferometers 
\cite {Kobayashi} 
as a Fano resonance tuning device.

\section{acknowledgments} M.  Mendoza would like to acknowledge the Brazilian agency CAPES for
financial support, while P.A.S.  is grateful to the continuous support provided by FAPESP.

\begin{figure} \caption{(a) Schematic illustration of the open quantum dot structure. 
 (b) Conductance
for the OQD for $w=3a$, continuos line, and $w=5$, dashed line. (c) LDOS for the energies labeled 
in (b), refering to the $w=3a$ case.}
\label{1} \end{figure}

\begin{figure} \caption{Total transmission probabilities (conductance) for the OQD as a 
function of incident electron energy, comparying the unperturbed with perturbed ones at the 
positions indicated in the insets. (a) dashed line for the $H=0.0$eV and continuous line 
for $H=0.2$eV. (b) $H=0.0$eV: dashed line. $H=0.5$eV: continuous line. (c) and (d): 
continuous line for $H=0.05$eV and dashed line for $H=4$ eV.}
\label{2} \end{figure}

\begin{figure} \caption{Conductance for the OQD as a function of incident electron energy for 
a perturbative spike at the center of the dot (dashed line) and near a corner (continuous line). For 
both cases $H=0.5$ eV.
} \label{3} \end{figure}

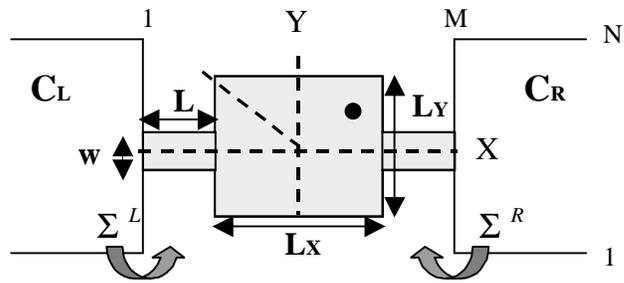
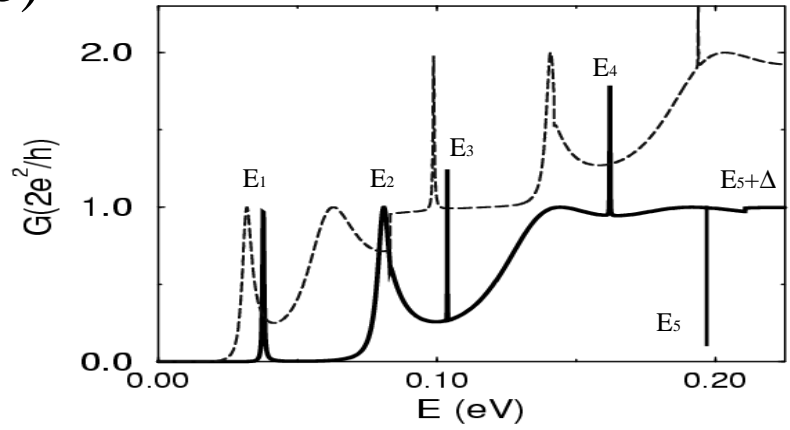
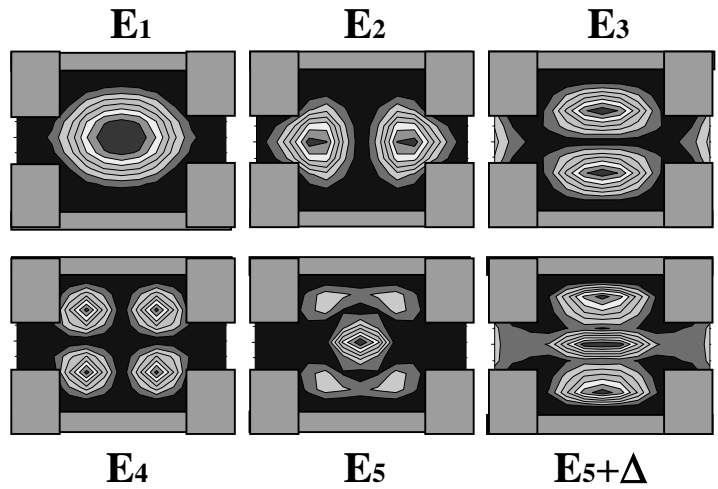
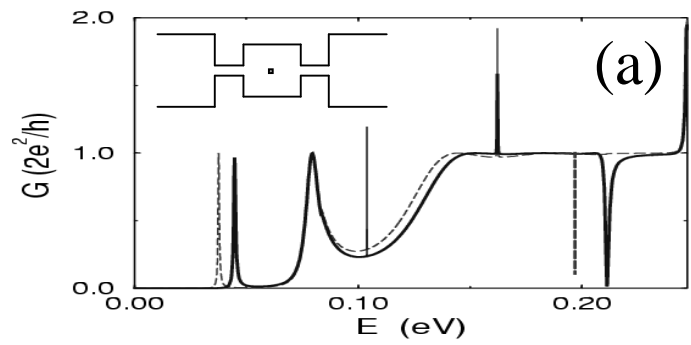
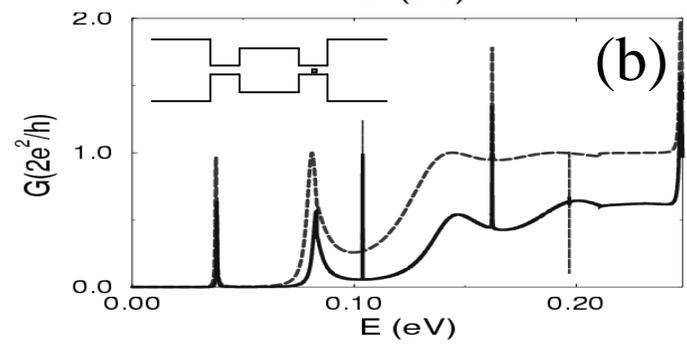
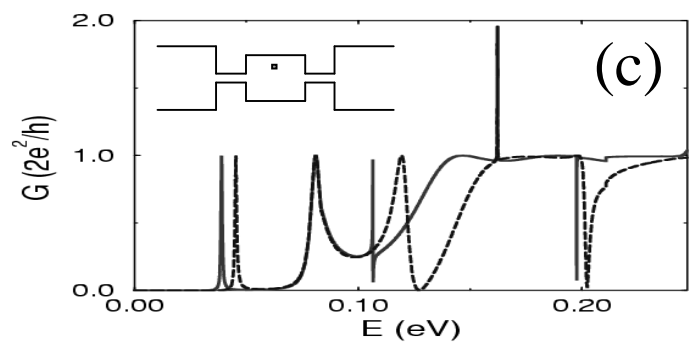
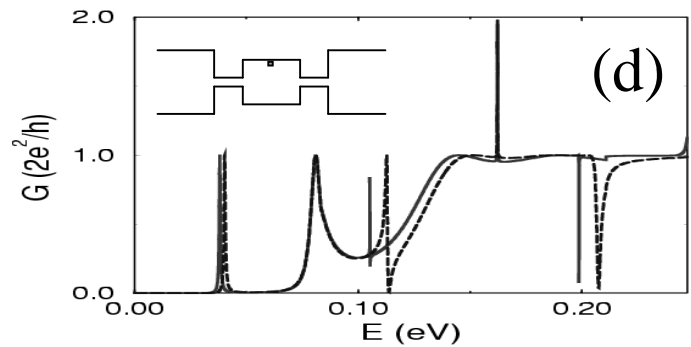
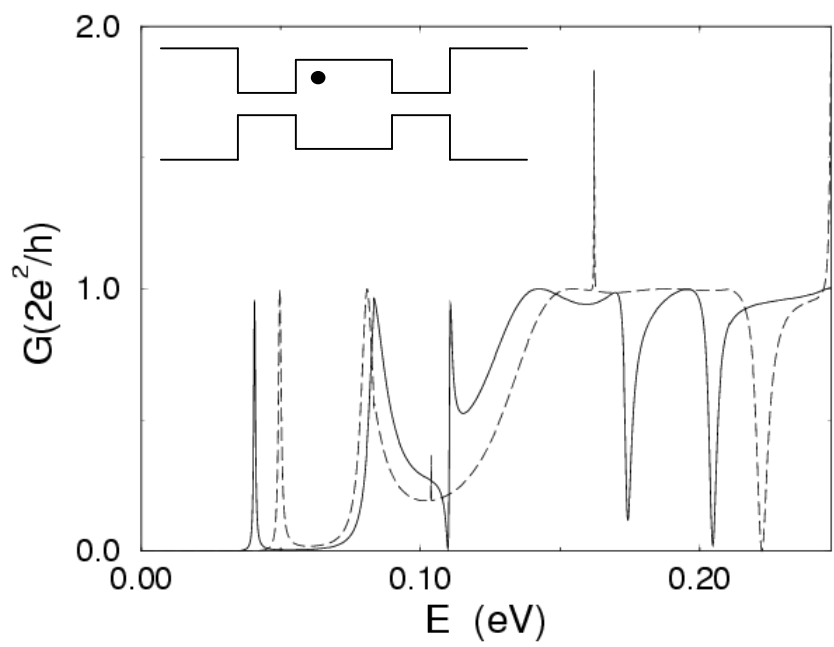
\end{document}